\documentclass[conference, 9pt]{IEEEtran}
\IEEEoverridecommandlockouts
\usepackage{cite}
\usepackage{amsmath,amssymb,amsfonts}
\usepackage{graphicx}
\usepackage{color, xcolor}
\usepackage{lipsum}
\usepackage{comment}
    \newcommand{\thicktilde}[1]{\mathbf{\tilde{\text{$#1$}}}}
\usepackage{tikz}
  \newlength\squareheight
\usepackage{multirow}
\usepackage{float}
\usepackage{subcaption}
\usepackage{siunitx}
\usepackage{oubraces}
\usepackage{algorithmic}
\usepackage{graphicx}
\usepackage{textcomp}
\usepackage{xcolor}
\usepackage[hang,flushmargin]{footmisc}
\usepackage[cal=boondoxupr]{mathalfa}
\def\BibTeX{{\rm B\kern-.05em{\sc i\kern-.025em b}\kern-.08em
    T\kern-.1667em\lower.7ex\hbox{E}\kern-.125emX}}

\usepackage{fancyhdr}
\usepackage{ragged2e}  % For \justifying
\fancypagestyle{firstpage}{
    \fancyhf{} % Clear all headers and footers
    \setlength{\footskip}{28pt}  % Reduce this value to move the footer up
    \fancyfoot[C]{\justifying \footnotesize{\copyright 2025 IEEE. Personal use of this material is permitted. However, permission to reprint/republish this material for advertising or promotional purposes or for creating new collective works for resale or redistribution to servers or lists, or to reuse any copyrighted component of this work in other works, must be obtained from the IEEE.}} % Footer text
}

\begin{document}

\title{F-StrIPE: Fast Structure-Informed Positional Encoding for Symbolic Music Generation\\
\thanks{This work was funded by the European Union (ERC, HI-Audio, 101052978). Views and opinions expressed are however those of the author(s) only and do not necessarily reflect those of the European Union or the European Research Council. Neither the European Union nor the granting authority can be held responsible for them. The accompanying website at \textit{bit.ly/faststructurepe} contains music samples and code.}
}

% \author{\IEEEauthorblockN{1\textsuperscript{st} Manvi Agarwal}
% \IEEEauthorblockA{
% \textit{T\'{e}l\'{e}com Paris}\\
% \textit{Institut Polytechnique de Paris}\\
% Paris, France \\
% \and
% \IEEEauthorblockN{2\textsuperscript{nd} Changhong Wang}
% \IEEEauthorblockA{
% \textit{T\'{e}l\'{e}com Paris}\\
% \textit{Institut Polytechnique de Paris}\\
% Paris, France \\
% \and
% \IEEEauthorblockN{3\textsuperscript{rd} Ga\"{e}l Richard}
% \IEEEauthorblockA{
% \textit{T\'{e}l\'{e}com Paris}\\
% \textit{Institut Polytechnique de Paris}\\
% Paris, France \\
% \author{\IEEEauthorblockN{Manvi Agarwal, %\IEEEauthorrefmark{1}, 
% Changhong Wang, %\IEEEauthorrefmark{2}, 
% Ga\"{e}l Richard%\IEEEauthorrefmark{3}, 
% %Author 4\IEEEauthorrefmark{3}, Author 5\IEEEauthorrefmark{4}
% }
% \IEEEauthorblockA{T\'{e}l\'{e}com Paris, Institut Polytechnique de Paris, France}
% }

\author{\IEEEauthorblockN{Manvi Agarwal, 
Changhong Wang, 
Ga\"{e}l Richard 
%Author 4\IEEEauthorrefmark{3}, Author 5\IEEEauthorrefmark{4}
}
\IEEEauthorblockA{LTCI, T\'{e}l\'{e}com Paris, Institut Polytechnique de Paris, France}
}

% \author{\IEEEauthorblockN{Michael Shell\IEEEauthorre fmark{1}, Homer Simpson\IEEEauthorrefmark{2}, James K irk\IEEEauthorrefmark{3}, Montgomery Scott\IEEEautho rrefmark{3} and Eldon Tyrell\IEEEauthorrefmark{4}} \IEEEauthorblockA{\IEEEauthorrefmark{1}School of Ele ctrical and Computer Engineering\\
% Georgia Institute of Technology, Atlanta, Georgia 30 332--0250\\
% Email: mshell@ece.gatech.edu} \IEEEauthorblockA{\IEEEauthorrefmark{2}Twentieth Cen tury Fox, Springfield, USA\\
% Email: homer@thesimpsons.com}
% \IEEEauthorblockA{\IEEEauthorrefmark{3}Starfleet Aca demy, San Francisco, California 96678-2391\\ Telephone: (800) 555--1212, Fax: (888) 555--1212} \IEEEauthorblockA{\IEEEauthorrefmark{4}Tyrell Inc.,
% 123 Replicant Street, Los Angeles, California 90210 --4321}}

\maketitle
\thispagestyle{firstpage}
\begin{abstract}
While music remains a challenging domain for generative models like Transformers, recent progress has been made by exploiting suitable musically-informed priors. One technique to leverage information about musical structure in Transformers is inserting such knowledge into the positional encoding (PE) module. However, Transformers carry a quadratic cost in sequence length. In this paper, we propose \emph{F-StrIPE}, a structure-informed PE scheme that works in linear complexity. Using existing kernel approximation techniques based on random features, we show that F-StrIPE is a generalization of Stochastic Positional Encoding (SPE). %\textit{We compare SPE and F-StrIPE from the perspective of musical structure, showing that our approach is appropriately grounded in both theory and domain knowledge. }
We illustrate the empirical merits of F-StrIPE using melody harmonization for symbolic music.
% \blfootnote{\copyright 2024 IEEE. \textit{Personal use of this material is permitted. Permission from IEEE must be obtained for all other uses, in any current or future media, including reprinting/republishing this material for advertising or promotional purposes, creating new collective works, for resale or redistribution to servers or lists, or reuse of any copyrighted component of this work in other works.}}
\end{abstract}

\begin{IEEEkeywords}
music generation, symbolic music, transformers, positional encoding, kernels.
\end{IEEEkeywords}

\section{Introduction}
\label{sec:intro}

% \begin{figure*}[h!]
%     \begin{subfigure}{2.05\columnwidth}
%     \centering
%     \includegraphics[scale=0.65]{figs/ismir_2024_sff.png}
%     % \caption{top: SVD 4x4 matrix}
%     % \label{subfig:a}
%     \end{subfigure}\\
%     \begin{subfigure}{2.05\columnwidth}
%     \centering
%     \includegraphics[scale=0.65]{figs/ismir_2024_rff.png}
%     % \caption{bottom: SVD 4x4 matrix calculation}
%     % \label{subfig:b}
%     \end{subfigure}
%     \caption{Approximating Method 3~\cite{huang_improve_2020} (left column, corresponding to Equation \ref{eq:huang_m3}) with two random feature methods (right column). The top row conceptually expresses Proposition 1, while the bottom row demonstrates how Random Fourier Features is connected to Stochastic Fourier Features. The variables referenced here are detailed in Section \ref{section:prelim}.}
%     \label{fig:visualize_matrices}
% \end{figure*}

Owing to their remarkable ability to produce realistic, high-quality samples, deep generative models are attracting significant interest. Two interdependent factors have been clearly established as being important for their superior performance: voluminous data and ever-increasing parameter counts~\cite{kaplan_scaling_2020}. Domains with abundant data -- text, vision and speech -- have successfully exploited these factors. In comparison, the limited size of publicly-available music datasets places an implicit limit on the size of the models that can be used, making music a challenging data domain. 

% Despite the advances brought about by the introduction of Transformers and attention, music generated by such models often lacks long-term coherence and organization~\cite{wu_jazz_2020}. 
The introduction of Transformers and attention has accelerated the advances in deep generative models and music has been no stranger to this phenomenon. However, generated music often lacks long-term coherence and organization, which are hallmarks of real music~\cite{wu_jazz_2020}.
% One way of improving music generation is to embed prior knowledge about musical structure into Transformers~\cite{richard_model_2024,bhandari_motifs_2024} through the positional encoding (PE) module~\cite{agarwal_structure_2024,yi_popmag_2020,guo_domain_2023,liu2022symphony}. 
One way of improving music generation is to embed prior knowledge about musical structure into data-driven models~\cite{ji_survey_2023,richard_model_2024,bhandari_motifs_2024}, for example, through the positional encoding (PE) module of Transformers~\cite{agarwal_structure_2024,yi_popmag_2020,guo_domain_2023,liu2022symphony}. 
This is an attractive option that provides a drop-in replacement for vanilla, structure-free PE without added complexity or training pipeline changes.

Despite their successes, Transformers bear a quadratic complexity in sequence length, which restricts their use on long sequences. Kernel approximations can be used to mitigate this cost~\cite{tay_efficient_2022, tsai_transformer_2019}.

\begin{figure}[t]
    \centering
    \includegraphics[width=1.0\linewidth]{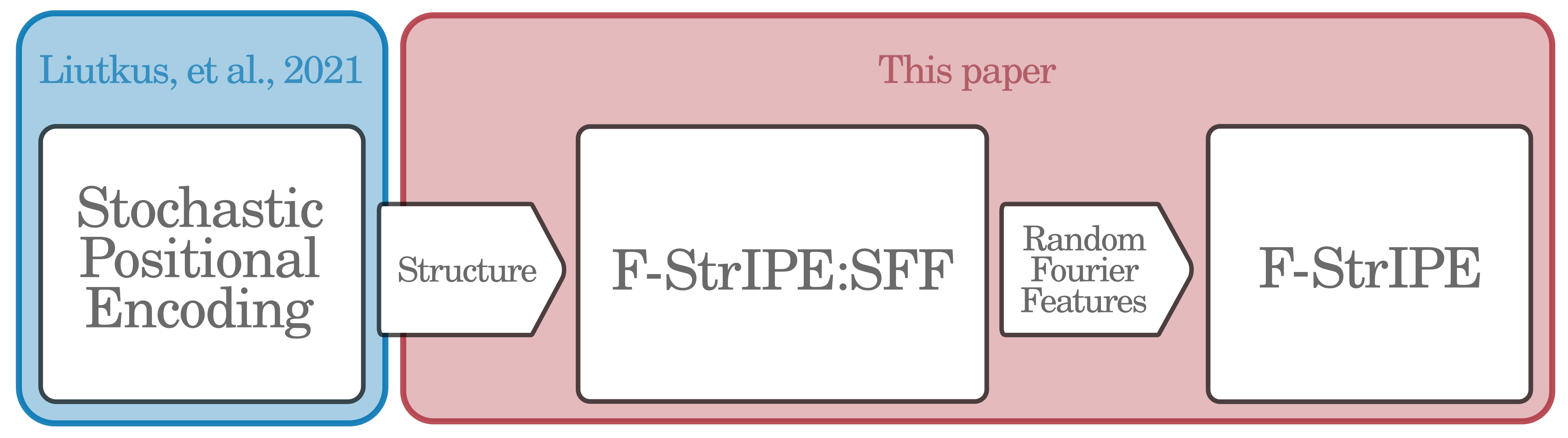}
    \caption{A schematic showing our main contributions (best viewed in colour).}
    \label{fig:contributions}
\end{figure}

In this work, we unite these two strands of research, one of which aims to improve Transformers for music generation by using informative priors, and the other that employs kernel approximations to achieve low-complexity Transformers that are able to process long sequences.
In particular, we propose F-StrIPE, a \textbf{f}ast \textbf{str}ucture-\textbf{i}nformed \textbf{p}ositional \textbf{e}ncoding method that works in linear complexity. We show that F-StrIPE is a generalization of Stochastic Positional Encoding (SPE)~\cite{liutkus_relative_2021}, an existing structure-free positional encoding technique, as sketched in Figure \ref{fig:contributions}. We do this by using Random Fourier Features~\cite{rahimi_random_2007}, thereby drawing on and providing a connection to previous work on kernel approximation for efficient attention. We empirically evaluate F-StrIPE on the symbolic music generation task of melody harmonization and show that, compared to the features used by SPE, Random Fourier Features are better-suited to structure-informed PE. We demonstrate how structure can be efficiently used in Transformers for music generation, giving us the coveted twin benefits of better performance and lower computational cost.
%\textit{In particular, we first generalize Stochastic Positional Encoding (SPE)~\cite{liutkus_relative_2021} to leverage structure. We draw on previous work on random features for kernel approximation and show that Random Fourier Features~\cite{rahimi_random_2007} is better for structure-informed PE than the features used by SPE. In this way, we demonstrate how structure can be efficiently used in Transformers for music generation, giving us the coveted twin benefits of better performance and lower computational cost.}

% \begin{figure*}[h!]
%     \centering
%     \includegraphics[width=1.6\columnwidth]{figs/R.png}
%     \caption{A visual representation of the connection between Stochastic Positional Encoding and Random Fourier Features. The variables referenced here are detailed in Sections \ref{section:background} and \ref{section:methods}.}
%     \label{fig:visualize_matrices}
% \end{figure*}

\begin{figure*}[h!]
    \centering
    \includegraphics[width=1.58\columnwidth]{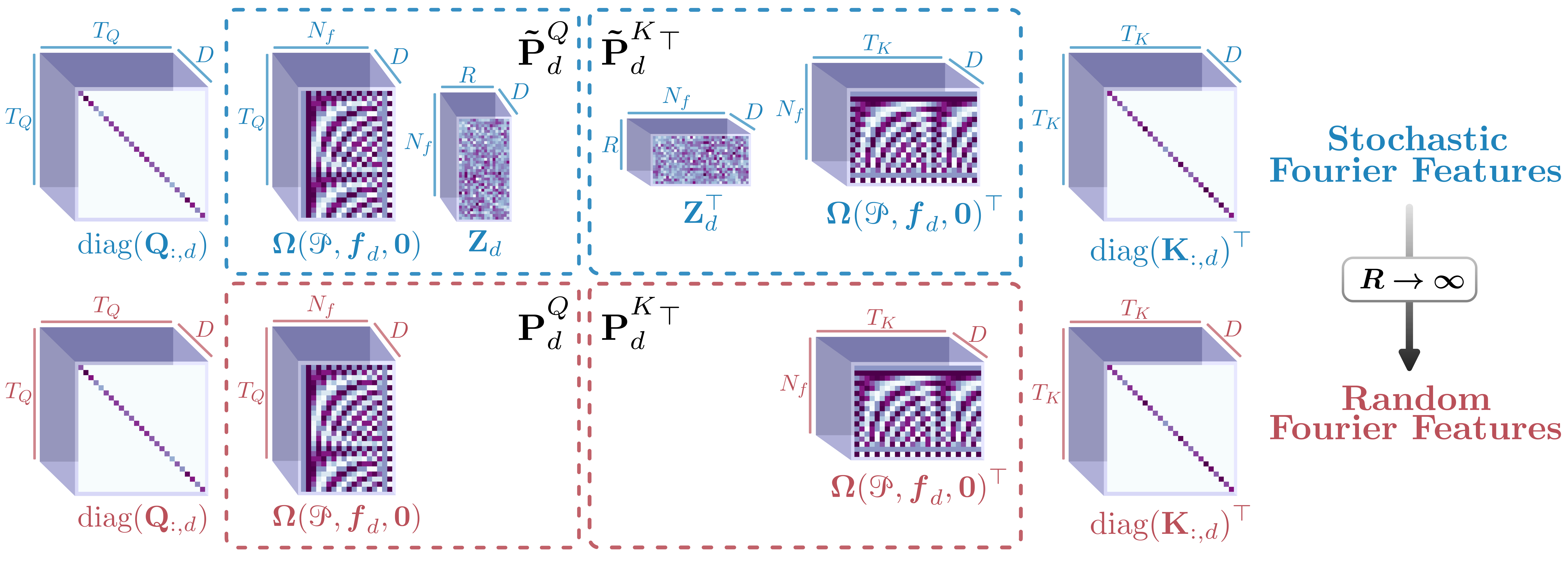}
    \caption{A visual representation of the connection between Stochastic Positional Encoding and Random Fourier Features (best viewed in colour). The variables referenced here are detailed in Sections \ref{section:background} and \ref{section:methods}.}
    \label{fig:visualize_matrices}
\end{figure*}

\section{Background}
\label{section:background}

\subsection{Transformers and Positional Encoding}

The Transformer~\cite{vaswani_attention_2017} is a sequence-to-sequence architecture that processes all timesteps in a sequence parallely using \textit{attention}. Given input sequence $[\mathbf{x}_1 , ... , \mathbf{x}_T]$, each element of the output sequence is:
\begin{equation}
    \mathbf{y}_m = \frac{\sum_n \mathsf{a}_{m n} \mathbf{v}_n}{\sum_n \mathsf{a}_{m n}} \text{ with } \mathsf{a}_{mn} = \text{exp} \bigg( \frac{a_{mn}}{\sqrt{D}} \bigg)
\end{equation}

%where $a_{mn} = \mathbf{q}_m \mathbf{k}_n^\top$ and $m, n \in \{1, ..., T\}$ are timesteps. 
where $a_{mn} = \mathbf{q}_m \mathbf{k}_n^\top$ is the attention coefficient or ``similarity score" for a pair of timesteps $(m,n)$, with $m, n \in \{1, ..., T\}$.
The query, key and value vectors $\mathbf{q}_m, \mathbf{k}_m, \mathbf{v}_m$ are obtained by linearly transforming input $\mathbf{x}_m$. Since attention executes pairwise computation among all timesteps, Transformers are invariant to permutations in the temporal order of inputs. Hence, positional encoding (PE) is applied to provide the model with a sense of time. Positional information can be incorporated in two places: at the input or during attention computation. The latter is called Relative Positional Encoding (RPE) and we focus on this approach here.

\subsection{Efficient Attention with Positional Information}

Attention has quadratic complexity in sequence length. To address this, a kernelized form of attention was introduced:
% \begin{align} 
%     \mathsf{a}_{mn} &= \mathcal{K} ( \mathbf{q}_m , \mathbf{k}_n ) = \mathbb{E}\left[ \phi(\mathbf{q}_m) \phi(\mathbf{k}_n)^{\top}\right] \label{eq:attention_as_expectation} \\
%     \boldsymbol{\mathsf{A}} &= \left[ \mathsf{a}_{mn} \right] \approx \phi(\mathbf{Q})\phi(\mathbf{K})^{\top} \label{eq:attention_approx}
% \end{align} 
\begin{equation} \label{eq:attention_approx}
    \mathsf{a}_{mn} = \mathcal{K} ( \mathbf{q}_m , \mathbf{k}_n ) = \mathbb{E} \Big[ \phi(\mathbf{q}_m) \phi(\mathbf{k}_n)^{\top} \Big]
\end{equation}

where $\mathcal{K}$ is a positive (semi)definite kernel and $\phi(\mathbf{x}): \mathbb{R}^D \to \mathbb{R}^{D_\phi}$ defines a \textit{randomized feature map} for $\mathbf{x}$~\cite{tsai_transformer_2019, choromanski_rethinking_2021}. With multiple instantiations, $\phi$ captures, on average, the relationship between $\mathbf{q}_m$ and $\mathbf{k}_n$, typified by $\mathcal{K}$. Thus, coefficients $\mathsf{a}_{mn}$ need not be computed explicitly, which produces linear-complexity Transformers.
%Because RPE requires explicit computation of $a_{mn}$, it was not possible to use RPE with the efficient formulation described above.

The efficient formulation described above is not directly applicable to RPE which, as introduced in \cite{shaw_rpe_2018}, requires the explicit computation of attention coefficients $a_{mn}$. This gap was addressed by Stochastic Positional Encoding~\cite{liutkus_relative_2021}, which approximates attention as:

\begin{equation} \label{eq:spe:ohyeah}
a_{mn} \approx \left[ \sum_{d=1}^D
    \lefteqn{\overbrace{\phantom{
    \text{diag}(\mathbf{Q}_{:,d})
    \frac{
        \thicktilde{\mathbf{P}}^Q_d
    }{\sqrt{R}}
    }}^{
    \mathbf{Q}^{\text{SFF}}
    }}
    \text{diag}(\mathbf{Q}_{:,d})
    \underbrace{
    \frac{
        \thicktilde{\mathbf{P}}^Q_d
    }{\sqrt{R}}
    \lefteqn{\overbrace{\phantom{
    \frac{
        % {\mathbf{P}_d^{K}}^\top
        \thicktilde{\mathbf{P}}^K_d{}^\top
    }{\sqrt{R}}
    \text{diag}(\mathbf{K}^\top_{:,d})
    }}^{
    \mathbf{K}^{\text{SFF}}
    }}
    \frac{
        \thicktilde{\mathbf{P}}^K_d{}^\top
    }{\sqrt{R}}
    }_{
    \approx \mathbf{P}_d
    }
    \text{diag}(\mathbf{K}^\top_{:,d})
    \right]_{mn}
\end{equation}

In SPE, content information ($\mathbf{Q}$/$\mathbf{K}$) is combined with context information ($\thicktilde{\mathbf{P}}^Q_d$/$\thicktilde{\mathbf{P}}^K_d$) in the matrices $\mathbf{Q}^{\text{SFF}}$ and $\mathbf{K}^{\text{SFF}}$. Note that, for the sake of brevity, we use the convention `$\mathbf{Q}$/$\mathbf{K}$' as a shorthand to mean `$\mathbf{Q}$ (respectively $\mathbf{K}$)' in this paper. SPE uses the feature maps $\phi$ proposed in earlier work~\cite{katharopoulos_transformers_2020,zhuoran_efficient_2021}, but applies them to $\mathbf{Q}^{\text{SFF}}$ and $\mathbf{K}^{\text{SFF}}$ (\ref{eq:spe:ohyeah}) instead of $\mathbf{Q}$ and $\mathbf{K}$ (\ref{eq:attention_approx}).
In (\ref{eq:spe:ohyeah}), $\mathbf{Q}_{:, d}/\mathbf{K}_{:, d}$ extracts a $T_Q/T_K$-dimensional vector containing the $d^{th}$ dimension for all timesteps of the query/key matrix. The positional matrix $\mathbf{P}_d$ captures the relationship between all pairs $(m, n)$ of timesteps that come from the positional index sequences $\mathcal{P}_Q = \{ 1, ..., m, ..., T_Q \}$ and $\mathcal{P}_K = \{ 1, ..., n, ..., T_K \}$. $\mathbf{P}_d$ is approximated by using $R$ feature realizations of $\mathcal{P}_Q$ and $\mathcal{P}_K$. These realizations are collected in the positional representation matrices $\thicktilde{\mathbf{P}}^Q_d$ and $\thicktilde{\mathbf{P}}^K_d$, which are given as:
\begin{equation} \label{eq:sff}
    \thicktilde{\mathbf{P}}^{Q/K}_d = \frac{\boldsymbol{\Omega}\left(\mathcal{P}_{Q/K}, \boldsymbol{f}_d, \boldsymbol{\theta}_d^{Q/K}\right) \text{diag}\left(\ddot{\boldsymbol{\lambda}_d}\right) \mathbf{Z}_d}{\sqrt{2 N_f}}
\end{equation}
The first term, $\boldsymbol{\Omega}\left(\mathcal{P}_{Q/K}, \boldsymbol{f}_d, \boldsymbol{\theta}_d^{Q/K}\right)$, represents sinusoidal features for the index sequence $\mathcal{P}_{Q/K}$. These features are parameterized by $N_f$ frequencies, collected in $\boldsymbol{f}_d$, and their phase shifts, collected in $\boldsymbol{\theta}_d$. The second term, $\text{diag}\left(\ddot{\boldsymbol{\lambda}_d}\right)$, consists of gains that apply to the sinusoidal features. Finally, the last term, $\mathbf{Z}_d$, consists of i.i.d. entries from a zero-mean, unit-variance Gaussian distribution. The sinusoidal features are given by:

% where $\mathbf{Z}_d$ consists of i.i.d. entries from a zero-mean, unit-variance Gaussian distribution and $\ddot{\boldsymbol{\lambda}_d}$ are gains that are applied to the sinusoidal feature matrix $\boldsymbol{\Omega}$. Phase-shifts $\boldsymbol{\theta}_d^K$ are set to $\boldsymbol{0}$, unlike $\boldsymbol{\theta}_d^Q$. The matrix $\boldsymbol{\Omega}$ represents a sequence of positions $\mathcal{P}$, parameterized by $N_f$ frequencies, collected in $\boldsymbol{f}$, and their phase shifts, collected in $\boldsymbol{\theta}$, as:

\begin{equation} \label{eq:sff:sine_features}
[\boldsymbol{\Omega}(\mathcal{P}, \boldsymbol{f}, \boldsymbol{\theta})]_{i j}= \begin{cases}\cos \left(2 \pi \mathbf{f}[\omega] p_i+\boldsymbol{\theta}[\omega]\right) & \text { if } j = 2\omega \\ \sin \left(2 \pi \mathbf{f}[\omega] p_i+\boldsymbol{\theta}[\omega] \right) & \text { else }\end{cases}
\end{equation}

where $i$ runs over timesteps, such that $p_i$ is the $i^{\text{th}}$ timestep of index sequence $\mathcal{P}$ and $\omega$ runs over frequencies. The variable $j$ allows us to apply the same frequency $\mathbf{f}[\omega]$ to a pair of cosine and sine features.

We call this feature representation technique, which uses (\ref{eq:sff}) and (\ref{eq:sff:sine_features}), \textit{Stochastic Fourier Features} (SFF). We visualize the different components of SFF in the first row of Figure \ref{fig:visualize_matrices}. In particular, we illustrate the features $\thicktilde{\mathbf{P}}^Q_d$ and $\thicktilde{\mathbf{P}}^K_d$ when all gains are 1 and all phase shifts are 0. 

\section{Methods} \label{section:methods}

\subsection{Structure-informed positional encoding}

As we hinted in Section \ref{section:background}, positional encoding depends on a sequence $\mathcal{P} = [ p_1, p_2, ... , p_T ]$ of positional indices. In standard PE, which does not utilize structure, $p_i = i$, which makes $\mathcal{P}$ a linear grid. When structure is included in PE, $p_i = s_\ell (i)$, where $s_\ell (i)$ gives the structural label at level $\ell$ (e.g. chord) for timestep $i$. In this case, $s_\ell (i) = s_\ell (i^{\prime})$ is possible for $i \neq i^{\prime}$, making $\mathcal{P}$ a non-linear grid. In fact, we can consider vectorial positional indices with multiple resolutions of structural organization. Consequently, we obtain $p_i = \mathbf{s} (i)$, where $\mathbf{s} (i) = [ s_1 (i) , ..., s_\ell (i), ..., s_L (i) ]$ is a vector of $L$ structural labels. Viewed in this way, PEs with structure can simply be used as a drop-in replacement for PEs without structure by replacing the form of $p_i$, giving us richer positional information. This is a way to flexibly represent domain-specific prior knowledge about the underlying data domain.

\subsection{Adding structure to SPE: F-StrIPE:SFF}

In order to use rich positional information in SPE, we can augment the sinusoidal feature matrix $\boldsymbol{\Omega}$ from (\ref{eq:sff:sine_features}) to be:
\begin{equation} \label{eq:pos_sine_features}
    [\boldsymbol{\Omega}(\mathcal{P}, \boldsymbol{f}, \boldsymbol{\theta})]_{i j}= \begin{cases}\cos \left(2 \pi \mathbf{f}[\omega, :]^\top p_i+\boldsymbol{\theta}[\omega]\right) & \text { if } j = 2\omega \\ \sin \left(2 \pi \mathbf{f}[\omega, :]^\top p_i+\boldsymbol{\theta}[\omega] \right) & \text { else }\end{cases}
\end{equation}
Here, we use the vectorial formulation of structure-aware positional indices $p_i = \mathbf{s}(i)$, unlike (\ref{eq:sff:sine_features}) where $p_i = i$ was a sequence of structure-free positional indices linked solely to the passage of time. Whereas in (\ref{eq:sff:sine_features}), $\mathbf{f}[\omega]$ was a single frequency, $\mathbf{f}[\omega, :]$ in (\ref{eq:pos_sine_features}) is a vector of frequencies. Therefore, each frequency $\mathbf{f}[\omega, \ell]$ in this vector acts on the $\ell^{\text{th}}$ structural label at timestep $i$. We can combine (\ref{eq:spe:ohyeah}), (\ref{eq:sff}) and (\ref{eq:pos_sine_features}) to obtain a fast structure-aware PE technique that uses Stochastic Fourier Features. We call this method \textit{F-StrIPE:SFF}.

% Here, unlike (\ref{eq:sff:sine_features}) where $\mathbf{f}[\omega]$ was a single frequency, $\mathbf{f}[\omega, :]$ is a vector of frequencies. Each frequency $\mathbf{f}[\omega, \ell]$ in this vector acts on the $\ell^{\text{th}}$ structural label at timestep $i$, where we can use the vectorial formulation of structure-aware positional indices $p_i = \mathbf{s}(i)$. In contrast, $p_i = i$ in (\ref{eq:sff}).

\subsection{Asymptotic case of SFF: Random Fourier Features} \label{ssection:asymptotic_rff}

Using Equations (\ref{eq:spe:ohyeah}) and (\ref{eq:sff}), we can express the SFF approximation of the positional matrix $\mathbf{P}_d$ for arbitrary timesteps $m$ and $n$ as:
\begin{equation} \label{eq:sff_mn}
    \mathbf{P}_d[m, n] \approx \Big[ \Omega_\mathcal{Q}^d [m, :] (\mathbf{Z}_d \mathbf{Z}_d^\top) \Omega_\mathcal{K}^{d^\top} [:, n] \Big] / R \\
\end{equation}

where we use the abbreviation $\Omega_\mathcal{A}^d = \boldsymbol{\Omega}\left(\mathcal{P}_A, \boldsymbol{f}_d, \boldsymbol{\theta}_d^A \right) \text{diag}\left(\ddot{\boldsymbol{\lambda}_d}\right)$. We observe that $\mathbf{Z}_d \mathbf{Z}_d^\top = \widehat{\mathbf{C}}_d$ acts as an empirical covariance matrix for the features $\Omega_\mathcal{Q}^d$ and $\Omega_\mathcal{K}^d$. Since $\mathbf{Z}_d$ has zero mean and unit variance, as $R \to \infty$, $\widehat{\mathbf{C}}_d$ approaches the theoretical covariance matrix $\mathbf{C}_d = \mathbf{I}_{2N_{f}}$.
In the ideal case of $\mathbf{C}_d$, (\ref{eq:sff_mn}) simplifies to $\mathbf{P}_d[m, n] \approx \Omega_\mathcal{Q}^d [m, :] \Omega_\mathcal{K}^{d^\top} [:, n]$, giving:
\begin{equation} \label{eq:ideal_C}
    \mathbf{P}_d[m, n] \approx  \frac{1}{N_f}  \sum_{\omega = 1}^{N_f} \Lambda_\omega \cos \Big( f_{\omega} ( \mathcal{P}_Q[m] - \mathcal{P}_K[n] ) + \Theta_\omega \Big)
\end{equation}
where $\Lambda_\omega$ is the gain contributed by the matrices $\text{diag}\left(\ddot{\boldsymbol{\lambda}_d}\right)$ and $\Theta_\omega$ is the phase-shift contributed by $\boldsymbol{\theta}^Q_d$ and $\boldsymbol{\theta}^K_d$. This representation has been studied in previous work, where it is called \textit{Random Fourier Features} (RFF) \cite{rahimi_random_2007, sutherland_error_2015}.

\subsection{Generalizing F-StrIPE:SFF to F-StrIPE}

Using this insight, we can redesign the positional feature matrices from (\ref{eq:sff}) to be:
\begin{equation} \label{eq:rff}
    \mathbf{P}^{Q/K}_d = \boldsymbol{\Omega}\left(\mathcal{P}_{Q/K}, \boldsymbol{f}_d, \boldsymbol{\theta}_d^{Q/K}\right) \text{diag}\left(\ddot{\boldsymbol{\lambda}_d}\right) / \sqrt{N_f}
\end{equation}
where the sinusoidal features $\boldsymbol{\Omega}$ uses structure-aware positional indices as given in (\ref{eq:pos_sine_features}). With this, we can now modify (\ref{eq:spe:ohyeah}) to use $\mathbf{P}^{Q/K}_d$ in place of $\thicktilde{\mathbf{P}}^{Q/K}_d$, giving us $\mathbf{Q}^{\text{RFF}}/\mathbf{K}^{\text{RFF}}$ in place of $\mathbf{Q}^{\text{SFF}}/\mathbf{K}^{\text{SFF}}$. To signify that such a PE technique generalizes F-StrIPE:SFF to use RFF in place of SFF, we call this method \textit{F-StrIPE}.

In the second row of Figure \ref{fig:visualize_matrices}, similar to SFF, we show the different components of RFF in the case where gains are 1 and phase shifts are 0. RFF can be understood as the ideal case of SFF where $R \to \infty$. Seen in this way, RFF gives us a noiseless estimate of $\mathbf{P}_d$ with direct access to the theoretical covariance matrix $\mathbf{C}_d$.

% In addition, if we pair $\mathbf{Q}^{\text{SFF}}/\mathbf{K}^{\text{SFF}}$ with the sinusoidal features given in Equation \ref{eq:sff:sinusoidal_features}, in place of those given in Equation \ref{eq:sff:sine_features}, we obtain a richer version of SPE which accepts multi-dimensional structural information instead of time indices. To distinguish it from SPE, we name this variant F-StrIPE:SFF.

\section{Experiments}
% To test the efficacy of our method, we use the task of melody harmonization for symbolic music. 
We assess the merits of our approaches on the task of melody harmonization for symbolic music.

\subsection{Dataset and Input Representation} \label{sssection:data_input}
We use the Chinese POP909 dataset~\cite{wang_pop909_2020} and three levels of structural labels with different resolutions~\cite{dai_automatic_2020}: melodic pitch ($16^{th}$-note), chord (quarter-note) and phrase (measure). Each MIDI file in this dataset consists of three tracks: melody, bridge (second melody) and piano (accompaniment). 
We use the POP909 alignment dataset~\cite{agarwal_structure_2024} to correctly match the structural labels with the input. We convert the MIDI files to binary pianorolls $\mathbf{X} \in \mathbb{B}^{(n_\text{tracks} \times 128) \times n_\text{time}}, \mathbb{B} = \{0, 1\}$, where $n_\text{tracks}$ is the number of tracks and $n_\text{time}$ is the number of timesteps in the pianoroll.

\subsection{Task Setup}
Given the sequence for the melody and bridge tracks $\big[ \mathbf{x}_n \in \mathbb{B}^{(n_\text{tracks} - 1) \times 128} \big]$ as input, with $n \in \{ 1, ..., n_\text{time}\}$, the model must predict all tracks $\big[ \mathbf{y}_n \in \mathbb{B}^{n_\text{tracks} \times 128} \big]$. We expect the model to produce the complete accompaniment track for all timesteps at once, without conditioning later predictions on earlier predictions. 
% We use three settings: (16, 16), (16, 64) and (64, 64), where the first number is the sequence length (in bars of music) used for training and the second number is that used for testing.
We use two settings: (16, 16) and (16, 64). The first number is the sequence length (in measures) for training and the second is that used for testing.

\subsection{Model and Training}
We use a 2-layer causal encoder Transformer with 4 heads and 512 model dimension. Training for 15 epochs with a batch size of 8, we use gradient clipping and curriculum learning~\cite{bengio_curriculum_2009}. We use two learning rate schedulers: a linear warmup and an epoch-wise decay.
We do a grid-search for two hyperparameters: learning rate (choices: $\{ 1, 5, 10 \} \times 0.0001$) and post-processing binarization strategy~\cite{agarwal_structure_2024} (choices: thresholding, thresholding with merge). While the first binarization strategy uses a fixed threshold, the second additionally fills the gap between notes if the gap is less than a minimum distance.

\subsection{Baselines and Our Methods} \label{sssection:baselines}
We consider three types of baselines: (i) Transformers without PE (NoPE~\cite{tsai_transformer_2019,haviv_transformer_2022}), (ii) Transformers with efficient, approximate atttention but no structural information in PE (SPE~\cite{liutkus_relative_2021}), and (iii) Transformers with structural information in PE but using inefficient, exact attention (S S-RPE~\cite{agarwal_structure_2024}).
From our methods, we use F-StrIPE with the three structural levels described in Section \ref{sssection:data_input}. We also assess the influence of different random features with F-StrIPE:SFF using all structural levels. 
We perform ablations on F-StrIPE by selecting one level at a time during training. Finally, we use the best-performing structural level from the F-StrIPE ablations to additionally do an ablation study with F-StrIPE:SFF.

\subsection{Evaluation}
We choose a collection of musically-motivated metrics from the literature, guided by four criteria. 

To assess large- and small-scale structural properties, we use Self-Similarity Matrix Distance (SSMD)~\cite{wu_musemorphose_2021}. For both the target and the prediction, we calculate chroma vectors, giving us the number of onset occurrences per chroma in every half-measure. We then compose a self-similarity matrix (SSM) for each chroma vector by taking the pairwise cosine similarities between all elements of the vector. The SSMD is the mean absolute difference between the SSM of the target and the SSM of the prediction.

For melodic consistency, we use Chroma Similarity (CS)~\cite{wu_musemorphose_2021}. Using the aforementioned method of constructing chroma vectors, we compute the CS as the mean cosine similarity between corresponding entries of the target chroma vector and the prediction chroma vector.

For rhythmic consistency, we use Grooving pattern Similarity (GS)~\cite{wu_jazz_2020}. The grooving pattern of a piece of music is a vector that encodes a 1 for the quarter-notes where onsets occur and 0 for those where no onsets occur. After obtaining the grooving patterns of the target and prediction, we compute the GS as the percentage of quarter-notes where the corresponding pattern values match.

To gauge polyphonicity, we use Note Density Distance (NDD)~\cite{agarwal_structure_2024, haki_real_2022}. We calculate the total number of pitches in each $16^{\text{th}}$-note of the target and prediction. The NDD is the average percentage of missing pitches in the prediction, with the number of pitches in the target giving us the maximum possible value.

% \input{sections/results_table_2decimals}
% \usepackage{multirow}
\begin{table*}[t]
\centering
\resizebox{2.05\columnwidth}{!}{%
\begin{tabular}{lcccccccc}
\hline
\multicolumn{1}{c|}{\multirow{3}{*}{\textbf{Method}}} & \multicolumn{4}{c|}{\textbf{Train = 16 bars; Test = 16 bars}} & \multicolumn{4}{c}{\textbf{Train = 16 bars; Test = 64 bars}} \\
\multicolumn{1}{c|}{} & \textbf{CS} & \textbf{SSMD} & \textbf{GS} & \multicolumn{1}{c|}{\textbf{NDD}} & \textbf{CS} & \textbf{SSMD} & \textbf{GS} & \multicolumn{1}{c}{\textbf{NDD}}\\
\multicolumn{1}{c|}{} & $\uparrow$ & $\downarrow$ & $\uparrow$ & \multicolumn{1}{c|}{$\downarrow$} & $\uparrow$ & $\downarrow$ & $\uparrow$ & \multicolumn{1}{c}{$\downarrow$} \\ \hline \hline
% \multicolumn{9}{c}{\textbf{Melody Harmonization}} \\ \hline \hline
\multicolumn{1}{l|}{\textbf{NoPE}} & \phantom{0}2.68 $\pm$ 0.2\phantom{*} & 29.31 $\pm$ 0.0* & \phantom{0}7.82 $\pm$ 0.1\phantom{*} & \multicolumn{1}{c|}{93.94 $\pm$ 0.0*} & \phantom{0}2.67 $\pm$ 0.2 & 27.60 $\pm$ 0.0* & \phantom{0}7.80 $\pm$ 0.2 & \multicolumn{1}{c}{92.49 $\pm$ 0.1\phantom{*}} \\
\multicolumn{1}{l|}{\textbf{S S-RPE}} & 14.52 $\pm$ 0.7\phantom{*} & 28.92 $\pm$ 0.1\phantom{*} & 21.58 $\pm$ 0.9\phantom{*} & \multicolumn{1}{c|}{88.48 $\pm$ 0.5\phantom{*}} & $\diagdown$ & $\diagdown$ & $\diagdown$ & \multicolumn{1}{c}{$\diagdown$} \\
\multicolumn{1}{l|}{\textbf{SPE} \cite{liutkus_relative_2021}} & \phantom{0}1.07 $\pm$ 0.0* & 29.33 $\pm$ 0.0* & \phantom{0}6.02 $\pm$ 0.0* & \multicolumn{1}{c|}{94.65 $\pm$ 0.0*} & \phantom{0}6.71 $\pm$ 0.3 & 27.59 $\pm$ 0.0* & 12.62 $\pm$ 0.3 & \multicolumn{1}{c}{91.48 $\pm$ 0.1\phantom{*}}\\ \hline
\multicolumn{1}{l|}{\textbf{F-StrIPE}}  & 11.84 $\pm$ 1.2\phantom{*} & 29.18 $\pm$ 0.0* & 18.62 $\pm$ 1.4\phantom{*} & \multicolumn{1}{c|}{90.93 $\pm$ 0.6\phantom{*}} & \phantom{0}9.13 $\pm$ 1.2 & 27.53 $\pm$ 0.0* & 14.70 $\pm$ 1.7 & \multicolumn{1}{c}{90.30 $\pm$ 0.3\phantom{*}} \\
\multicolumn{1}{l|}{\textbf{F-StrIPE:M}} & \phantom{00}1.9 $\pm$ 0.1\phantom{*} & 29.31 $\pm$ 0.0* & \phantom{0}7.07 $\pm$ 0.1\phantom{*} & \multicolumn{1}{c|}{94.27 $\pm$ 0.0*} & \phantom{0}2.22 $\pm$ 0.1 & 27.60 $\pm$ 0.0* & \phantom{0}7.42 $\pm$ 0.2 & \multicolumn{1}{c}{92.69 $\pm$ 0.0*} \\
\multicolumn{1}{l|}{\textbf{F-StrIPE:C}} & \textbf{16.61 $\pm$ 1.5\phantom{*}} & \textbf{28.71 $\pm$ 0.1\phantom{*}} & \textbf{23.19 $\pm$ 2.7\phantom{*}} & \multicolumn{1}{c|}{\textbf{86.42 $\pm$ 0.4\phantom{*}}} & \textbf{13.29 $\pm$ 1.1} & \textbf{27.15 $\pm$ 0.1\phantom{*}} & \textbf{16.73 $\pm$ 0.9} & \multicolumn{1}{c}{\textbf{85.61 $\pm$ 0.4\phantom{*}}} \\
\multicolumn{1}{l|}{\textbf{F-StrIPE:P}} & \phantom{0}2.07 $\pm$ 0.1\phantom{*} & 29.31 $\pm$ 0.0* & \phantom{0}7.23 $\pm$ 0.1\phantom{*} & \multicolumn{1}{c|}{94.20 $\pm$ 0.0*} & \phantom{0}2.31 $\pm$ 0.1 & 27.60 $\pm$ 0.0* & \phantom{0}7.56 $\pm$ 0.1 & \multicolumn{1}{c}{92.62 $\pm$ 0.0*} \\ \hline
\multicolumn{1}{l|}{\textbf{F-StrIPE:SFF}} & \phantom{0}2.75 $\pm$ 3.0\phantom{*} & 29.32 $\pm$ 0.0* & \phantom{0}8.15 $\pm$ 3.5\phantom{*} & \multicolumn{1}{c|}{94.18 $\pm$ 0.9\phantom{*}} & \phantom{0}6.49 $\pm$ 0.3 & 27.58 $\pm$ 0.0* & 12.09 $\pm$ 0.7 & \multicolumn{1}{c}{91.46 $\pm$ 0.1\phantom{*}} \\ %\hline
\multicolumn{1}{l|}{\textbf{F-StrIPE:SFF:C}} & \phantom{0}4.72 $\pm$ 3.9\phantom{*} & 29.25 $\pm$ 0.1\phantom{*} & 10.22 $\pm$ 4.6\phantom{*} & \multicolumn{1}{c|}{93.11 $\pm$ 1.5\phantom{*}} & 10.35 $\pm$ 1.0 & 27.44 $\pm$ 0.1\phantom{*} & 12.09 $\pm$ 0.7 & \multicolumn{1}{c}{89.32 $\pm$ 0.6\phantom{*}} \\ \hline
\end{tabular}
}
\caption{Performance on melody harmonization. F-StrIPE:(M/C/P) are the ablations on F-StrIPE, described in Section \ref{sssection:baselines}, that apply RFF on only one structural level at a time - melodic pitch/chord/phrase. F-StrIPE:SFF:C applies SFF on chords, which is the best performing ablation setting with RFF. `0.0*' refers to standard deviations that are lower than 0.05. `$\diagdown$' refers to simulations where the given inference setting could not be accessed due to the heavy computational demands of the method.}
\label{tab:results_unrounded}
\end{table*}

\section{Results and Discussion}

\begin{table}[b]
\centering
\resizebox{\columnwidth}{!}{%
\begin{tabular}{ccc}
\hline
\textbf{Method} & \textbf{Additional Parameters} & \textbf{Runtime Space Complexity} \\ \hline \hline
S S-RPE~\cite{agarwal_structure_2024}  &  $\mathcal{O} ( \mathcal{s} (\mathcal{h} \mathcal{d})^2 + \ell (\mathcal{h} \mathcal{d})^2 )$  &  $\mathcal{O}( \ell \mathcal{t}^2 \mathcal{h} \mathcal{d} )$  \\
SPE~\cite{liutkus_relative_2021}  & $\mathcal{O} (\mathcal{h} \mathcal{d} N_f)$ & $\mathcal{O}(\ell \mathcal{t} \mathcal{h} \mathcal{d} N_f)$ \\ 
F-StrIPE & $\mathcal{O} (\mathcal{s} \mathcal{h} \mathcal{d} N_f)$ & $\mathcal{O}(\mathcal{s} \ell \mathcal{t} \mathcal{h} \mathcal{d} N_f)$ \\ \hline                         
\end{tabular}%
}
\caption{Complexity analysis for different methods with $\ell$ layers, $\mathcal{h}$ heads, $\mathcal{d}$ head dimension, $\mathcal{s}$ structures, and $\mathcal{t}$ sequence length. A typical order for size is $\mathcal{s} < (\mathcal{h}, \ell) << \mathcal{d} <<< \mathcal{t} $, with $\mathcal{s}$ contributing the least and $\mathcal{t}$ contributing the most, assuming equal growth rate.}
\label{tab:complexity}
\end{table}

In Table \ref{tab:results_unrounded}, we report the mean and standard deviation on 3 seeds for each metric.

\subsection{Utility of structural information in PE}

When we compare PEs without structure (NoPE, SPE) against PEs with structure (S S-RPE, F-StrIPE), we observe that the latter perform better. This matches previous findings reported in the literature~\cite{agarwal_structure_2024} which argued that using structure in PE boosts performance, particularly in underdetermined problems such as melody harmonization.

\subsection{Influence of different random features}

F-StrIPE:SFF adds structural information to SPE and F-StrIPE improves on this by using RFF in place of SFF. F-StrIPE:SFF yields marginal improvements over the performance of SPE in the (16, 16) scenario, in particular, on CS. This lends some additional support to our previous observation that structural information used in PE is useful. However, F-StrIPE, which uses a noise-free estimate of the positional matrix $\mathbf{P}_d$, gives us significant boosts over the performance of F-StrIPE:SFF and, by extension, that of SPE, on both (16, 16) and (16, 64). These improvements are particularly noticeable in CS, GS and NDD and are especially pronounced in the (16, 16) setting. This emphasizes that the correct approximation techniques can strongly enhance the effect of augmenting our models with prior knowledge.

\subsection{Ablations on F-StrIPE and F-StrIPE:SFF}

As described in Section \ref{sssection:baselines}, we perform ablations on structural levels used during training, resulting in models that are specialized to use only melody (F-StrIPE:M), only chord (F-StrIPE:C) or only phrase (F-StrIPE:P) as structural information. When we compare these models against F-StrIPE, which uses all three structures, we see first that F-StrIPE:M and F-StrIPE:P do worse than F-StrIPE on all metrics and both task settings. In fact, their performance drops lower than even NoPE. In contrast, F-StrIPE:C brings a clear advantage over F-StrIPE, with significant improvements on all metrics and both task settings. This fits our intuition that the accompaniment in pop songs can be nicely characterized by chord progressions~\cite{paiement_probabilistic_2005,zhu_xiaoice_2018}. Our results show that using only chord information is better than using all structures simultaneously. This shows that while task-specific structural information can boost performance, ill-founded and generic priors can prove counterproductive. Thus, how prior knowledge is selected and incorporated into a deep-learning model should be an important consideration while designing such systems.

\subsection{Comparing complexities}

These results should be viewed in the context of the complexity analysis presented in Table \ref{tab:complexity}. Compared to S S-RPE, SPE and F-StrIPE have linear complexity in sequence length, which makes a sizeable dent in the requirement for computational resources. Moreover, scaling up from SPE to F-StrIPE only adds a factor of $\mathcal{s}$, corresponding to the number of structures we use in PE, which grows the slowest compared to all other variables. In fact, since the ablations use only a single structure at a time $(\mathcal{s} = 1)$, the complexity of our best-performing method, F-StrIPE:C, matches that of SPE.

Thus, on the one hand, in the worst case where multiple structures are needed, it only costs a small amount of additional computational resources. On the other hand, if we already know which structure is best-suited to our task, we can benefit from F-StrIPE and leverage prior knowledge in our model without needing any extra resources.

\subsection{Length Generalization}

Finally, we see that models that are trained on 16 bars of music but tested on 64 bars of music reflect the same trends as seen in the models that were tested on 16 bars of music: structural information combined with RFF provides a sizeable improvement over baselines that either do not use structure or use structure but with SFF.

On CS and GS, the PEs that use SFF (SPE, F-StrIPE:SFF and F-StrIPE:SFF:C) show large improvements on the (16, 64) setting compared to the (16, 16) setting. This does not hold true for the RFF-based PEs, where some methods show small improvements and others show small deteriorations. Nevertheless, F-StrIPE:C outperforms all other PEs in the (16, 64) situation. This suggests that an in-depth comparative investigation of the characteristics of different approximation techniques is needed. Specifically, it would be interesting to understand which approximation is suited to what learning scenario and whether we can combine the strengths of different approximations to obtain a more robust, fast, structure-informed PE. Interestingly, in two metrics - SSMD and NDD - all the methods in Table \ref{tab:results_unrounded} do marginally better in the (16, 64) scenario compared to the (16, 16) one.

The transfer of performance from the (16, 16) setting to the (16, 64) setting can be partly attributed to the presence of stereotypical structure and a high degree of repetition in pop songs~\cite{sargent_estimating_2017,dai_missing_2022}. Thus, 16 measures of music could potentially contain much of the necessary information to generate much longer sequences. It would be interesting to quantatively assess whether this hypothesis is true.
% To assess this hypothesis, we first assumed that an $\mathcal{N}$-measure sample in our dataset consists of $(n_\text{tracks} \times 128) \times |\mathcal{N}|$ independent variables, where $|\mathcal{N}|$ is the number of timesteps in $\mathcal{N}$ measures of music at half-beat resolution following our convention in Section \ref{sssection:data_input}. We plot the average Shannon entropy over all such variables, with the distribution for a given variable being computed over the full dataset. The results are plotted in Fig. \ref{fig:entropy}. We see that samples with 16 bars of music have the same informational content as samples with 64 bars of music. %Although a more detailed investigation is needed to understand this phenomenon, ...

% \begin{figure}[h!]
%     \centering
%     \includegraphics[width=0.5\linewidth]{figs/entropy.png}
%     \caption{entropy}
%     \label{fig:entropy}
% \end{figure}
% \textbf{Appropriate to point this out? Doesn't highlight our method in particular even though it does the best out of all metrics.}

\section{Conclusion}

In this paper, we demonstrated how structural information can be used in linear-complexity positional encoding, thereby retaining superior performance without sacrificing efficiency. We did this by first extending SPE to accept multi-resolution, structure-aware positional indices, obtaining F-StrIPE:SFF. Then, we showed the connection between SPE~\cite{liutkus_relative_2021} and Random Fourier Features~\cite{rahimi_random_2007} and developed a novel method, called F-StrIPE, framed as a structure-aware generalization of SPE. The combination of these two interventions --- using structure and using Random Fourier Features --- gave us a fast, structure-informed positional encoding method that outperformed SPE, F-StrIPE:SFF and other competitive baselines on melody harmonization for symbolic music.

% We did this by first showing the connection between SPE~\cite{liutkus_relative_2021} and Random Fourier Features~\cite{rahimi_random_2007}. We, then, demonstrated how we can include multi-resolution structures within such efficient PE methods that use kernel approximation techniques. , which we called F-StrIPE. We exhibited the empirical validity of F-StrIPE using a symbolic music generation task.

% M3~\cite{huang_improve_2020} and SPE~\cite{liutkus_relative_2021} compute the same form of PE. We, then, extended these methods to include multi-resolution structures by establishing that Random Fourier Features provides a better approximation than SPE-style Stochastic Fourier Features. We proved the validity of our structure-informed PE methods using both synthetic protocols grounded in theory and real-world experiments with two music generation tasks.

\bibliographystyle{./IEEEbib/IEEEtran}
\bibliography{strings}

\end{document}